\documentclass[aps,twocolumn,prl,showpacs]{revtex4}

\usepackage{graphics}
\usepackage{epsfig}
\usepackage{graphicx}
\usepackage{amsmath,amssymb,amsthm,bbm,latexsym}    
\usepackage{amsthm} 
\usepackage{amsbsy} 

\newcommand{\ndash}{\textendash}
\newcommand{\mdash}{\textemdash}
\newcommand{\ie}{{\it i.e.\,}}

\newcommand{\etal}{{\it et al.\,}}

\newcommand{\me}{\mathrm{e}}
\newcommand{\mi}{\mathrm{i}}
\newcommand{\md}{\mathrm{d}}

\newcommand{\tr}{\mathrm{tr}\,} 
\newcommand{\bra}[1]{\left\langle#1\right|} 
\newcommand{\ket}[1]{\left|#1\right\rangle} 
\newcommand{\overlap}[2]{\left\langle #1 \, \vrule \,#2 \right\rangle}

\newcommand{\bracket}[3]{\left\langle #1 \,\vrule\, #2 \,\vrule\, #3 \right\rangle}

\newcommand{\ucl}{\affiliation{Department of Physics and Astronomy, University College London, 
Gower Street, London WC1E 6BT, United Kingdom}}

\begin{document}
\title{Single-copy entanglement in a gapped quantum spin chain}
\author{Christopher Hadley}\ucl
\pacs{75.10.Pq,	03.65.Ud, 03.67.-a}
\date{\today}

\begin{abstract}
The {\it single-copy entanglement} of a given many-body system is defined [J. Eisert and M. Cramer, Phys. Rev. A. {\bf 72}, 042112 (2005)] as the maximal entanglement deterministically distillable from a bipartition of a single specimen of that system.  For critical (gapless) spin chains, it was recently shown that this is exactly {\it half} the von Neumann entropy [R. Or\'us, J. I. Latorre, J. Eisert, and M. Cramer, Phys. Rev. A {\bf 73}, 060303(R) (2006)], itself defined as the entanglement distillable in the asymptotic limit---\ie given an infinite number of copies of the system.  It is an open question as to what the equivalent behaviour for gapped systems is.  In this paper, I show that for the paradigmatic spin-$S$ Affleck--Kennedy--Lieb--Tasaki chain (the archetypal gapped chain), the single-copy entanglement is {\it equal} to the von Neumann entropy: \ie all the entanglement present may be distilled from a single specimen.  
\end{abstract}
\maketitle



The amount of entanglement naturally present in the ground states of many-body quantum systems is of great interest to both the quantum information and condensed matter communities \cite{article2007amico-fazio-osterloh-vedral}.  For bipartite pure states the unique asymptotically continuous entanglement monotone is the von Neumann entropy \cite{article1996bennett-bernstein-popescu-schumacher, article1997rohrlich-popescu}; the amount of entanglement present in the ground states of spin chains and other many-body sytems is often quantified by calculating this quantity with respect to a particular bipartition.  It is asymptotically equal to the ratio of the number of EPR singlets one may distill from a given system to the number of copies of that system\mdash that is, given an infinite number of copies of the system \cite{article1996bennett-bernstein-popescu-schumacher}.  Of course in realistic, physical situations one might only have access to a single specimen of a particular system, and the von Neumann entropy in this case gives an upper bound to the distillable entanglement.  An interesting question then arising is: how much entanglement can be deterministically distilled from a single copy of a system?  

This was studied originally by Lo and Popescu \cite{article2001lo-popescu}, who considered various optimal strategies for the interconversion of entangled states by local operations.  It was subsequently shown by Nielsen \cite{article1999nielsen} that deterministic transformation of one state $\ket{\psi}$ to another $\ket{\phi}$ using local operations and classical communication is possible if the nonincreasingly ordered Schmidt coefficients of the final state majorize those of the initial state ($\lambda_\psi \prec \lambda_\phi$); this was then generalised to a strategy for converting arbitrary bipartite states by Vidal \cite{article1999vidal, article1999vidal-2}.  

More recently, Eisert and Cramer \cite{article2005eisert-cramer} and Or\'us \etal \cite{article2006orus-latorre-eisert-cramer} defined the {\it single-copy entanglement} as the number of singlets deterministically distillable from a single specimen of a given system, and showed that for the case of gapless quantum spin chains close to criticality this is exactly half the von Neumann entropy: that is, {\it half the entanglement present may be distilled in a single process} \footnote{The single-copy entanglement has also been studied recently by Peschel and Zhao \cite{article2005peschel-zhao}, and Zhou \etal \cite{article2006zhou-barthel-fjaerestad-schollwock}.}. However, a large class of spin chains of interest are {\it gapped} (\ie there is an energy gap between the ground and first excited states), which have substantially different entanglement properties: in the gapless case, the {\it block entropy} (the entanglement of a block of contiguous spins with the remainder) is known to be logarithmically divergent in the block length, and in the gapped case, it saturates \cite{article2003vidal-latorre-rico-kitaev, article2004keating-mezzadri, article2004jin-korepin-2, article2004calabrese-cardy, article2005keating-mezzadri}.  It is this short-ranged nature of entanglement in gapped systems which gives rise to area laws \cite{article2005plenio-eisert-dreissig-cramer, article2006cramer-eisert-plenio-dreissig, article2008wolf-verstraete-hastings-cirac, article2007hastings}.  It would therefore be interesting to know whether gapped chains also exhibit different behaviour in the context of single specimens.  While this remains an open problem in general, I shall in this paper provide an example giving indications that this is indeed the case.  Specifically, I shall show that the celebrated spin-$S$ Affleck\ndash Kennedy\ndash Lieb\ndash Tasaki (AKLT) chain \cite{article1987affleck-kennedy-lieb-tasaki, article1988affleck-kennedy-lieb-tasaki, article1989affleck} has a single-copy entanglement {\it equal} to the von Neumann entropy for all $S$; \ie {\it all the entanglement present in the spin chain may be distilled in a single process}.  This spin chain is a paradigm in condensed matter physics, and was the first evidence of the veracity of the Haldane conjecture \cite{article1983haldane-1, article1983haldane-2}.

Recent years have seen a resurgence of interest in this state, partly due to its role as the simplest of the matrix product states, which in general have been shown to efficiently simulate many 1D systems \cite{article1995totsuka-suzuki, article2003vidal, article2004vidal, article2007perezgarcia-verstraete-wolf-cirac} and may be used as a variational set in density matrix renormalisation group calculations \cite{article2005schollwoeck,article2005verstraete-cirac-latorre-rico-wolf}.  It has also attracted interest in the context of quantum information theory, due to its interesting entanglement properties \cite{article2004fan-korepin-roychowdhury, article2007fan-korepin-roychowdhury-hadley-bose, article2007katsura-hirano-hatsugai}, and the fact that all stabilizer states (including cluster states) have an interpretation in terms of a VBS state \cite{article2004verstraete-cirac, article2006clark}.

{\it Single-copy entanglement}.---The single-copy entanglement (with respect to a particular bipartition) is defined \cite{article2005eisert-cramer} as the maximal number of singlets one can deterministically distill from a single-copy of a specimen in a single process; \ie the single-copy entanglement is $E_1 = \log M$, if $M$ is the largest $m$ for which the transformation 
\begin{align}
\rho \to \ket{\psi_m}\bra{\psi_m}
\end{align}
is possible under local operations and classical communication with unit probability (where $\ket{\psi_m} = 1/\sqrt{m} \sum_{i=1}^m \ket{i,i}$, the maximally entangled state of dimension $m\times m$).  This is the equivalent number of singlets \footnote{All logarithms are taken to base 2.}.  Applying Nielsen's majorization criterion to a bipartition of a spin-$S$ chain into a block of length $L$ and the remainder, this holds if and only if 
\begin{align}
\sum_{k=1}^K\alpha^{\downarrow}_k \le \frac{K}{M}\,\,{\rm for\,\,all}\,\,K\in [1,M], 
\end{align}
where $\{ \alpha^\downarrow_1 , \ldots ,\alpha^\downarrow_{(2S+1)^L} \}$ are the nonincreasing eigenvalues of the reduced density matrix of the block of length $L$.  As pointed out by Eisert and Cramer  \cite{article2005eisert-cramer}, this is obviously equivalent to the criterion that $\alpha^\downarrow_1\le 1/M$ (\ie the majorization reduction must be more mixed than the maximally entangled state).  One can then define the single-copy entanglement
\begin{align}
E_1 (\rho) = \log  \left[ 1 / \alpha^\downarrow_1 \right ] = -\log \alpha^\downarrow_1 .
\end{align}

{\it Critical, gapless systems}.---Using the machinery of conformal field theory, Or\'us \etal found that for all translationally invariant quantum spin systems that can be mapped onto an isotropic, quadratic system of fermions (via the Jordan\ndash Wigner transformation), the single-copy entanglement of a block of length $L$ is exactly half the von Neumann entropy in the thermodynamic ($L\to\infty$) limit: both logarithmically diverge with $L$.  Explicity, 
\begin{align}
E_1 (\rho_L) = \frac{c}{6}\ln L - \frac{c}{6} \frac{\pi^2}{\ln L} + O(1/L),
\end{align}
where $c$ is the conformal field theoretic central charge.  

{\it Noncritical, gapped systems}.---It is known that for gapped systems, the block entropy saturates to a constant bound \cite{article2003vidal-latorre-rico-kitaev, article2004fan-korepin-roychowdhury, article2007katsura-hirano-hatsugai}.  This qualitatively different entropic behaviour would suggest that the single-copy entanglement might also have substantially different behaviour.  As a first step towards understanding this, it would be instructive to consider the AKLT chain, the first example of a chain satisfying the Haldane conjecture \cite{article1987affleck-kennedy-lieb-tasaki, article1988affleck-kennedy-lieb-tasaki, article1989affleck}.  This chain consists of $N$ spins of magnitude $S$, with two spin-$S/2$'s at either end.  In the Schwinger boson representation this is written \cite{article1988arovas-auerbach-haldane}
\begin{align}
\ket{{\rm VBS}} = \prod_{i=0}^N (a^\dag_i b^\dag_{i+1} - b^\dag_i a^\dag_{i+1})^S\ket{0},\label{aklt-boson}
\end{align}
where $a^\dag_i$, $b^\dag_i$ are bosonic operators, and the spin operators are defined as $S_i^+ = a^\dag_i b_i$, $S_i^- = a_i b_i^\dag$, $S_i^z = (a_i^\dag a_i - b^\dag_i b_i )/2$, with the constraint $a_i^\dag a_i + b_i^\dag b_i = 2S$.  For general $S$, this is the unique ground state of the Hamiltonian
\begin{align}
H = \sum_{j=1}^{N-1}\sum_{J=S+1}^{2S}A_J P_{j,j+1}^J + \pi_{0,1} + \pi_{N,N+1}, \label{aklt-spin}
\end{align}
where the operator $P_{j,j+1}$ projects bond spins $j$, $j+1$ onto the (symmetric) subspace of total spin $J$, and the $A_J$ are arbitrary positive coefficients.  The boundary terms $\pi_{0,1}$, $\pi_{N,N+1}$ are similarly defined to project the end spin $S$ and $S/2$ onto the total spin $J$ subspace.  The isotropic case with no boundary operators has also been studied for the case $S=1$ \cite{article2007fan-korepin-roychowdhury-hadley-bose}, and the edge effects decay exponentially; similarly the effect of studying periodic boundary conditions in Eq. (\ref{aklt-spin}) as opposed to open boundary conditions also decays exponentially \cite{article1991freitag-muellerhartmann}.
This state may be interpreted in terms of `bonds' between spins (Fig. \ref{diagram}), with each bond being a singlet, and is often referred to as a valence bond solid.

\begin{small}\begin{figure}
 \begin{center}\begin{tabular}{c}
   \includegraphics[width=9cm,angle=0]{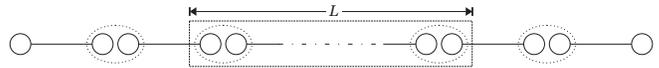}
  \end{tabular}\end{center}
 \caption{The VBS state has an interpretation in terms of `bonds' between spins, where each bond is a singlet of two spin-$S/2$'s, and at each bulk spin the state is projected (dotted circles) to the symmetric space of total spin.   The single-copy entanglement measured here is with respect to the partitioning into a block of length $L$ and the remainder.}
 \label{diagram}
\end{figure}\end{small}

{\it Coherent spin state approach to the VBS}.---An alternative representation of state (\ref{aklt-boson}) is given in terms of coherent spin states.  The well-known oscillator coherent state is defined to be an eigenstate of the annihilation operator $a$; by analogy, the coherent spin state is defined as an eigenstate of the spin raising operator $S^+$ \cite{article1971radcliffe}
\begin{align}
\ket{\theta, \phi} = \frac{1}{(1+|\mu |^2)^{2S}}\exp(\mu S^-)\ket{S,S},
\end{align}
where $\ket{S,m}$ is the spin state with $\langle {\bf S}^2 \rangle = S(S+1)$ and $\langle S_z \rangle = m$.  Parametrizing this with $\mu = \me^{\mi\phi}\tan(\theta /2)$ gives
\begin{align}
\ket{\theta, \phi} = \sum_{m=-S}^S\ u^{S+m}v^{S-m} \sqrt{{2S\choose S+m}} \ket{S,m},\label{coherent}
\end{align}
where $(u,v):=(\me^{\mi\phi/2}\cos(\theta /2), \me^{-\mi\phi/2}\sin(\theta / 2))$.  This state has a clear geometric interpretation: the state $\ket{\theta ,\phi}$ may be represented by the unit vector ${\bf \Omega} = (\theta,\phi)$ (\ie a point on the unit sphere).  Therefore the overlap between two such states may be found geometrically as $
\overlap{\theta ,\phi}{\theta',\phi'} = [ \cos(\theta /2)\cos(\theta' /2) + \sin(\theta /2) \sin(\theta' /2)\exp(\mi (\phi - \phi'))]^{2S}$, and thus
\begin{align}
|\overlap{\theta ,\phi}{\theta' ,\phi'}| = \left( \frac{1+{\bf \Omega\cdot \Omega}'}{2}\right)^S.
\end{align}
Comparing Eq. (\ref{coherent}) to the Schwinger boson representation of the ${\bf S}^2$, $S_z$ eigenstates
\begin{align}
\ket{S,m} = \frac{(a^\dag)^{S+m}(b^\dag)^{S-m}}{\sqrt{(S+m)!(S-m)!}}\ket{0},
\end{align}
it is clear that one can obtain an alternate representation of a state by making the replacement $a^\dag \to u$, $b^\dag \to v$, $a\to\partial /\partial u$, $b\to\partial /\partial v$, and multiplying by $\sqrt{(2S)!}$ when operators occur in pairs.  

Therefore, denoting the coherent state $\ket{\Omega}$ one may write \cite{article1988arovas-auerbach-haldane, article1991freitag-muellerhartmann} $\overlap{\Omega}{\rm VBS} = \prod_{i=0}^N \sqrt{(2S)!} (u_i v_{i+1} - v_i u_{i+1})^S$; and hence $|\overlap{\Omega}{\rm VBS}|^2 = \prod_{i=0}^N (2S)! |u_i v_{i+1} - v_i u_{i+1}|^{2S} =\prod_{\langle ij\rangle} (2S)!(1-{\bf\Omega}_k\cdot{\bf\Omega}_{k+1})^S/2^S.$
This approach was used in Ref. \cite{article1991freitag-muellerhartmann} to calculate all two-spin correlation functions, and more recently in Ref. \cite{article2007katsura-hirano-hatsugai} to calculate the block entropy.  


{\it Single-copy entanglement of the VBS}.---Using the coherent spin state approach, one may calculate the density matrix $\rho_L$ of $L$ contiguous spins. The trace of any operator in this representation is $\tr \mathcal{A} = (2S+1)/4\pi\int\md\Omega\bracket{\Omega}{\mathcal{A}}{\Omega}$ and therefore one obtains
\begin{align}
\rho_L &= \int\prod_{j\notin L} \frac{\md\Omega}{4\pi} \frac{\overlap{\Omega_j}{\rm{VBS}} \overlap{{\rm VBS}}{\Omega_j}}{\overlap{{\rm VBS}}{{\rm VBS}}}\nonumber\\
&= \frac{\int \prod_{i=1}^L\frac{\md\Omega_i}{4\pi}\prod_{k=1}^{L-1} T_{k,k+1}   \ket{\Omega_1}_0\bra{\Omega_1}\otimes\ket{\Omega_L}_{L+1}\bra{\Omega_L}}{\int \prod_{j=1}^L\frac{\md\Omega_j}{4\pi}\prod_{k=1}^{L-1} T_{k,k+1}},\nonumber
\end{align}
where the transfer matrix $T_{k,k+1}:=(1-{\bf\Omega}_k\cdot{\bf\Omega}_{k+1})^S/2^S$ and in the first line I have omitted numerical factors.  It was found in Ref. \cite{article2007katsura-hirano-hatsugai} that this is independent of the length of the total chain $N$, and therefore without loss of generality one can set $L=N$.  Following the methods of Refs. \cite{article1991freitag-muellerhartmann, article2007katsura-hirano-hatsugai}, one may find the eigenvalues of this matrix (replacing $S\to S/2$ for the end spins)

The following decomposition in terms of Legendre polynomials may be used \cite{article1991freitag-muellerhartmann, book1980gradshteyn-rhyzhik}
\begin{align}
\left(\frac{1+x}{2}\right)^S = \sum_{l=0}^S (2l+1) \frac{S!S!}{(S-l)!(S+l+1)!} P_l(x),
\end{align}
from whence it follows that
\begin{align}
\rho_L = \frac{4\pi}{(S+1)^2} \sum_{l=0}^S \lambda (l)^{L-1} I_l({\bf s}_0\cdot {\bf s}_{L+1}),
\end{align}
where $\lambda (l) := (-)^l S!(S+1)!/ (S-l)!(S+l+1)!$ and $I_l(X)$ is an $l$th order polynomial in $X$ determined recursively through 
\begin{align}
I_{j+1}(X) = \frac{2j+3}{(S+j+2)^2}\left(\frac{4X}{j+1}+j\right) I_j(X)\nonumber\\
 - \frac{j}{j+1}\frac{2j+3}{2j-1}\left(\frac{S-j+1}{S+j+2}\right)^2 I_{j-1}(X)
\end{align}
with $I_0(X)=1/4\pi$, $I_1(X) = 3X/4\pi (S/2+1)^2$.  These polynomials form a complete set of isotropic, two-site tensor operators.  

The eigenvalues of the density matrix may be found using this method, and summed using the standard formula for the von Neumann entropy [$S(\rho)=-\tr\rho\log\rho$] to find the block entropy for general $S$, which was found to approach $2\log (S+1)$ exponentially fast in $L$ (the thermodynamic limit) \cite{article2007katsura-hirano-hatsugai}, confirming the conjecture by Vidal \etal that the block entropy of a gapped integer spin chain reaches saturation for all $S$ \cite{article2003vidal-latorre-rico-kitaev}.  For the purposes of this paper, only the {\it largest} eigenvalue is required.  

The density matrix is diagonal in the basis of the total spin of spins $0$ and $L+1$.  These spins, of course, add up to several multiplets in the usual manner of spin addition, and so there will be degeneracy in the eigenvalues.  One thus requires the largest value of 
\begin{align}
\langle P_\sigma \rangle = \tr  \left\{ P_\sigma\rho_L \right\},
\end{align}
where $P_\sigma$ is the projector on to the subspace of total spin $\sigma$.  This multiplet distribution (\ie the eigenvalues multiplied by their weight) is given by \cite{article1991freitag-muellerhartmann}
\begin{align}
\langle P_\sigma \rangle = 4\pi (2\sigma+1)\sum_{j=0}^S  \frac{(S+j+1)!(S-j)!}{(S+1)!(S+1)!}\lambda (j)^{L+1} I_j[X(\sigma)], 
\end{align}
where $X(\sigma):= {\bf s}_0\cdot {\bf s}_{L+1} = \sigma (\sigma +1)/2 - S/2(S/2+1)$.  The calculation of the single-copy entanglement only requires the eigenvalues, and thus one omits the weights $(2\sigma +1)$.  

This distribution is found recursively, determined by the coefficients $I_j[X(\sigma)]$.  The largest values are given by the case $\sigma = S$ ($\sigma = 0$) for $L$ even (odd).  For even $L$, the required value is
\begin{align}
\langle P_{S} \rangle = (2S+1)\sum_{j=0}^S \frac{2j+1}{(S+1)^2} \lambda (j)^{L+1}
\end{align}
and thus the largest eigenvalue is
\begin{align}
\Lambda_1 &= \sum_{j=0}^S \frac{2j+1}{(S+1)^2} \lambda(j)^{L+1}\\
&= \frac{1}{(S+1)^2} \left\{ 1+\sum_{j=1}^S \lambda (j)^{L+1} (2j+1) \right\},
\end{align}
which gives the single-copy entanglement:
\begin{align}
&E_1 = -\log \Lambda_1 \nonumber\\
&= 2\log (S+1) -\log \left\{ 1+\sum_{j=1}^S \lambda (j)^{L+1} (2j+1) \right\}\\
&= 2\log (S+1) -\log \lambda (0)^{L+1} \left\{ 1+\sum_{j=1}^S \frac{\lambda (j)^{L+1}}{\lambda (0)^{L+1}} (2j+1) \right\}.
\end{align}
Since $\lambda (j) > \lambda (j+1)$ for all $j$, and $\lambda (0)=1$, it is clear that in the thermodynamic case $L\to\infty$ (as considered in the critical, gapless case \cite{article2006orus-latorre-eisert-cramer}), this becomes
\begin{align}
E_1 \to 2\log (S+1),
\end{align}
which is exactly equal to the von Neumann entropy, as found by Katsura \etal \cite{article2007katsura-hirano-hatsugai}.  The proof for $L$ odd follows analogously.  It therefore is the case that {\it all} the entanglement present in the VBS state (the ground state of the gapped spin-$S$ AKLT Hamiltonian) may be distilled from a single copy: one can distill with certainty a maximally entangled state, the dimension of which is related to $S$.  This would appear to have an intuitive explanation in terms of the valence bond picture of the state: the entanglement between a block and the remainder of the chain is related to the number of bonds `cut' by the boundary (indeed, this is similar to the reasoning behind area laws \cite{article2005plenio-eisert-dreissig-cramer, article2006cramer-eisert-plenio-dreissig, article2008wolf-verstraete-hastings-cirac, article2007hastings} although the analogy is not strict in this case, since the entanglements of formation and distillation are only equal in the asymptotic limit), and is further evidence of the qualitatively different behaviour of gapped chains to gapless chains.  One should contrast this with the critical case \cite{article2006orus-latorre-eisert-cramer}, from whence one can distill with certainty (in the $L\to\infty$ limit) a maximally entangled state of arbitrary dimension (\ie an infinite single-copy entanglement); the crucial, qualitative difference is that this is still only {\it half} the total amount of entanglement present.


{\it Summary}.---It has been demonstrated that all the entanglement present in the VBS ground state of the gapped AKLT Hamiltonian for arbitrary $S$ may be distilled with certainty in a single process.  This qualitative difference from the behaviour of gapless, critical chains provides evidence that the entanglement present in gapped systems is of a fundamentally different nature.  An open problem is whether the single-copy entanglement saturates to the von Neuman entropy for all gapped quantum systems.


I acknowledge financial support from UK EPSRC grants EP/P500559/1 and GR/S62796/01, and the University of London Valerie Myerscough Fund.  I thank V. E. Korepin for helpful discussions, and for his hospitality during a recent visit to the C. N. Yang Institute for Theoretical Physics at SUNY Stony Brook; and Sougato Bose, Dan Browne and Alessio Serafini for reading and commenting on the manuscript.  


\end{document}